\documentclass[12pt]{article}

\usepackage{amsmath,amssymb,amscd,amsbsy,amsgen,amsopn,amstext,
amsxtra}

\usepackage[dvips]{graphicx}

\begin{document}

\begin{flushright}
\end{flushright}

\vskip 0.5 truecm

\begin{center}
{\Large{\bf  Geometric phases for mixed states and decoherence}}
\end{center}
\vskip .5 truecm
\centerline{\bf  Kazuo Fujikawa }
\vskip .4 truecm
\centerline {\it Institute of Quantum Science, College of 
Science and Technology}
\centerline {\it Nihon University, Chiyoda-ku, Tokyo 101-8308, 
Japan}
\vskip 0.5 truecm

\makeatletter
\@addtoreset{equation}{section}
\def\theequation{\thesection.\arabic{equation}}
\makeatother

\begin{abstract}
The gauge invariance of geometric phases for mixed states is 
analyzed by using the hidden local gauge symmetry which arises 
from the arbitrariness of the choice of the basis set 
defining the coordinates in the functional space. This 
approach  gives a reformulation of the past results of adiabatic,
non-adiabatic and mixed state geometric phases.  
The geometric phases are identified uniquely as the holonomy associated with the hidden local gauge symmetry which is an 
exact symmetry of the Schr\"{o}dinger equation. The purification and its inverse in the description of de-coherent mixed states are consistent with the hidden local gauge symmetry. A salient feature of the present  formulation is that the total phase and visibility in the mixed state, which are directly observable in the interference 
experiment, are manifestly gauge invariant.
\end{abstract}


\section{Introduction}
The phase is a fundamental notion in quantum 
mechanics\cite{dirac,streater}, and the study of geometric 
phases is an attempt to understand quantum mechanics better. 
The notion of gauge symmetry or equivalence class plays a basic
role in the analysis of geometric phases\cite{pancharatnam,
mead, berry,simon,wilczek,anandan, aharonov, berry2, anandan2,
samuel,bohm,aitchison,mukunda,garcia,mostafazadeh,gauge-ind}, 
but its 
origin in the theory without gauge fields was not clear.
It has been recently shown that the gauge symmetry without gauge
 fields is identified with the hidden local gauge symmetry 
inherent in the second quantization, namely, the 
arbitrariness of the phase choice of  coordinates in the 
functional 
space~\cite{fujikawa, fujikawa2}. To be specific, one has the 
expansion of the field operator
\begin{eqnarray}
\hat{\psi}(t,\vec{x})=\sum_{n}\hat{b}_{n}(t)v_{n}(t,\vec{x})
\end{eqnarray}
which is invariant under the simultaneous 
transformations, $\hat{b}_{n}(t)\rightarrow
e^{-i\alpha_{n}(t)}\hat{b}_{n}(t)$ and $v_{n}(t,\vec{x})
\rightarrow e^{i\alpha_{n}(t)}v_{n}(t,\vec{x})$. 
For the generic case, we thus have the local symmetry 
$U=U(1)\times U(1)\times U(1) ...$, and in the presence of 
the degeneracy of basis vectors we have 
$U=U(n_{1})\times U(n_{2})\times U(n_{3}) ...$\cite{fujikawa}.
Any formulation
should preserve this exact symmetry, and the basic observation is
that the Schr\"{o}dinger amplitude with 
$\psi_{n}(0,\vec{x})=v_{n}(0,\vec{x})$ is transformed under this
 symmetry as $\psi_{n}(t,\vec{x})\rightarrow e^{i\alpha_{n}(0)}
\psi_{n}(t,\vec{x})$ independently of $t$ and thus the 
Schr\"{o}dinger equation is invariant under this symmetry.
It has been shown that both of the adiabatic\cite{fujikawa} and 
non-adiabatic\cite{fujikawa3} phases are uniquely specified as 
the holonomy associated with this symmetry, and the adiabatic
and non-adiabatic phases are treated on a completely equal 
footing. This formulation emphasizes that the geometric 
phases arise from the holonomy of the basis vectors rather than 
from the holonomy of the Schr\"{o}dinger amplitudes. The basic
aspects of the hidden local symmetry are summarized in Appendix 
A.
  
The definitions of geometric phases in mixed states have been
proposed by several authors~\cite{uhlmann, sjoqvist}, and it 
appears that the direct 
connection of the geometric phase with the 
geometrical picture of the motion of the polarization vector, 
for example, is lost. The gauge symmetries and associated 
holonomy in the 
geometric phases for mixed states have also been 
discussed~\cite{uhlmann, sjoqvist,slater,ericsson,
singh}, but the quantities invariant under  
the gauge symmetries discussed so far are non-linear 
(to be precise, not bi-linear) in the 
Schr\"{o}dinger amplitudes, and the total phase and visibility 
which are directly observable in the interference 
experiment~\cite{sjoqvist} are not invariant under the gauge 
symmetries discussed so far.

In the present paper, we analyze the gauge invariance of 
geometric phases for mixed states from the point of view of the 
hidden gauge symmetry explained above for pure states. We discuss
the geometric phases for mixed states proposed 
in~\cite{sjoqvist},
 which are direct generalizations of geometric phases in pure 
states. The geometric phases thus defined are known to be an 
intrinsic property of the mixed states without referring
to the ancilla subsystem~\cite{slater,ericsson}. On the other 
hand, the geometric phases proposed in~\cite{uhlmann}
 are based on the ``purification'' and crucially depend on 
the ancilla part~\cite{slater,ericsson}; these geometric phases 
may have some implications in quantum information, for example, 
but we forgo their analysis. 
Our basic observation is that an exact 
treatment of the Schr\"{o}dinger equation automatically contains
 all the geometric phases which are uniquely identified as the 
holonomy associated with  the 
hidden local symmetry, without going through an 
analysis of the projective Hilbert space.  More importantly, 
the total phase and visibility which are directly observable in 
the interference experiment~\cite{sjoqvist} are manifestly 
invariant under the hidden local gauge symmetry.

\section{Geometric phases in mixed states}

\subsection{Conventional formulation}

In the conventional formulation of geometric phases it is 
customary to classify the geometric phases into adiabatic and 
non-adiabatic ones. In connection with the geometric phase for
mixed states, we first briefly summarize the customary analysis 
of non-adiabatic phase which starts with~\cite{aharonov}
\begin{eqnarray} 
&&i\hbar\partial_{t}\psi(t,\vec{x})=\hat{H}(t)\psi(t,\vec{x}),
\ \ \ 
\int d^{3}x \psi^{\dagger}(t,\vec{x})\psi(t,\vec{x})=1
\end{eqnarray}
and the cyclic evolution is defined by
\begin{eqnarray}
\psi(T,\vec{x})=e^{i\phi}\psi(0,\vec{x})
\end{eqnarray}
with a constant $\phi$.

The equivalence class of state vectors, i.e.,``projective Hilbert
space'', is defined by~\cite{aharonov, samuel} 
\begin{eqnarray}
\{e^{i\alpha(t)}\psi(t,\vec{x})\},
\end{eqnarray}
and  we have an equivalence class of Hamiltonians
\begin{eqnarray}
\{\hat{H}(t)-\hbar\partial\alpha(t)\}
\end{eqnarray}
with an arbitrary function $\alpha(t)$ to maintain the 
Schr\"{o}dinger equation.
The total phase of the cyclic evolution is given by
\begin{eqnarray}
\phi_{T}=\rm arg \psi^{\dagger}(0,\vec{x})\psi(T,\vec{x})=
\rm arg \psi^{\dagger}(0,\vec{x}){\cal U}(t)\psi(0,\vec{x})
\end{eqnarray}
with
\begin{eqnarray}
{\cal U}(t)=T^{\star}
\exp[-\frac{i}{\hbar}\int_{0}^{t} \hat{H}(t)dt]
\end{eqnarray}
where $T^{\star}$ stands for the time ordering operation.
Note that the total phase is not invariant under the 
equivalence class (2.3).
One may then define the ``dynamical phase'' by
\begin{eqnarray}
\phi_{D}&=&-\frac{1}{\hbar}\int_{0}^{T}dt \int d^{3}x
\psi^{\dagger}(t,\vec{x})\hat{H}(t)\psi(t,\vec{x})
\nonumber\\
&=&-i\int_{0}^{T}dt \int d^{3}x
\psi^{\dagger}(t,\vec{x})\partial_{t}\psi(t,\vec{x}).
\end{eqnarray}
The non-adiabatic phase defined by the difference of 
$\phi_{T}$ and $\phi_{D}$
\begin{eqnarray}
\phi_{G}&=&{\rm arg} \psi^{\dagger}(0,\vec{x})\psi(T,\vec{x})
+\frac{1}{\hbar}\int_{0}^{T}dt \int d^{3}x
\psi^{\dagger}(t,\vec{x})\hat{H}(t)\psi(t,\vec{x})\nonumber\\
&=& {\rm arg} \psi^{\dagger}(0,\vec{x}){\cal U}(t)\psi(0,\vec{x})
+i\int_{0}^{T}dt \int d^{3}x
\psi^{\dagger}(0,\vec{x}){\cal U}(t)^{\dagger}\dot{{\cal U}}(t)
\psi(0,\vec{x})
\end{eqnarray}
is invariant under the equivalence class (2.3) and (2.4). This 
is customarily called the gauge invariance 
condition of non-adiabatic phase.

By using the equivalence class, one may choose a representative
$\bar{\psi}(t,\vec{x})=e^{i\alpha(t)}\psi(t,\vec{x})$ to satisfy
the parallel transport condition
\begin{eqnarray}
i\int d^{3}x
\bar{\psi}^{\dagger}(t,\vec{x})\partial_{t}\bar{\psi}(t,\vec{x})
=i\int d^{3}x
\psi^{\dagger}(t,\vec{x})\partial_{t}\psi(t,\vec{x})-\partial_{t}
\alpha(t)=0
\end{eqnarray}
namely,
\begin{eqnarray}
\bar{\psi}(t,\vec{x})=\exp{[i\int_{0}^{t}dt\int d^{3}x
\psi^{\dagger}(t,\vec{x})i\partial_{t}\psi(t,\vec{x})]}
\psi(t,\vec{x}).
\end{eqnarray}
For this $\bar{\psi}(t,\vec{x})$, which is non-linear in the 
Schr\"{o}dinger amplitude, the non-adiabatic phase for 
the pure state is given by~\cite{aitchison, mukunda}
\begin{eqnarray}
\phi_{G}&=&{\rm arg} \bar{\psi}^{\dagger}(0,\vec{x})
\bar{\psi}(T,\vec{x})\nonumber\\
&=&{\rm arg} \{\psi^{\dagger}(0,\vec{x})
\exp{[i\int_{0}^{T}dt\int d^{3}x
\psi^{\dagger}(t,\vec{x})i\partial_{t}\psi(t,\vec{x})]}
\psi(T,\vec{x}) \}.
\end{eqnarray}
This quantity is confirmed to be invariant under the gauge 
transformation, and $\phi_{G}$ stands for the holonomy of the 
vector $\bar{\psi}(t,\vec{x})$ for  cyclic evolution. 

In the case of the mixed state described by the density matrix
$\rho(t)={\cal U}(t)\rho(0){\cal U}(t)^{\dagger}$, the 
interference pattern is given by~\cite{sjoqvist} 
\begin{eqnarray}
I\propto 1+ |{\rm Tr}{\cal U}(T)\rho(0)|\cos[\chi
-{\rm arg Tr}{\cal U}(T)\rho(0)]
\end{eqnarray}
where $\chi$ stands for the variable $U(1)$ phase in one of the 
interference beams. The total phase of the mixed state is thus
given by 
\begin{eqnarray}
\gamma_{T}={\rm arg Tr}{\cal U}(T)\rho(0).
\end{eqnarray}
In a basis where the density matrix is written as 
\begin{eqnarray}
\rho(0)=\sum_{k=1}^{N}\omega_{k}|k\rangle\langle k|
\end{eqnarray}
the total phase is written as
\begin{eqnarray}
\gamma_{T}=\sum_{k=1}^{N}\omega_{k}{\rm arg}
[\langle k|{\cal U}(T)|k\rangle].
\end{eqnarray}
The parallel transport condition for a mixed state is imposed by
requiring that ${\rm Tr}\rho(t){\cal U}(t+dt)
{\cal U}(t)^{\dagger}$ be real and positive which in turn leads 
to
\begin{eqnarray}
{\rm Tr}\rho(t)\dot{{\cal U}}(t){\cal U}(t)^{\dagger}=
{\rm Tr}\rho(0){\cal U}(t)^{\dagger}\dot{{\cal U}}(t)=0
\end{eqnarray}
or equivalently
\begin{eqnarray}
\sum_{k=1}^{N}\omega_{k}[\langle k|{\cal U}(t)^{\dagger}
\dot{{\cal U}}(t)|k\rangle]=0.
\end{eqnarray}
Under this condition, the ``dynamical phase'' defined by 
\begin{eqnarray}
\gamma_{D}=-\frac{1}{\hbar}\int_{0}^{T}dt 
{\rm Tr}[\rho(t)\hat{H}(t)]=-i\int_{0}^{T}dt
{\rm Tr}[\rho(0){\cal U}(t)^{\dagger}\dot{{\cal U}}(t)]
\end{eqnarray}
vanishes identically.
It has also been asserted~\cite{sjoqvist} that the condition 
(2.16) or (2.17) while necessary, is not sufficient. Instead, 
the stronger conditions~\cite{wagh}
\begin{eqnarray}
\langle k|{\cal U}(t)^{\dagger}\dot{{\cal U}}(t)|k\rangle=0,
 \ \ \ k=1,2, .. N
\end{eqnarray}
have been proposed; all the constituent pure states in the mixed
state are required to be parallel transported independently.

To understand the prescription (2.19), it is convenient to 
analyze the N-state density matrix in the diagonal 
basis. On may
 then consider the transformation~\cite{singh}
\begin{eqnarray}
{\cal U}(t)\rightarrow {\cal U}^{\prime}(t)={\cal U}(t)
\sum_{k}^{N}e^{i\theta_{k}(t)}|k\rangle\langle k|
\end{eqnarray}
then the orbit of the density matrix remains unchanged
\begin{eqnarray}
\rho(0)\rightarrow \rho^{\prime}(t)=
{\cal U}^{\prime}(t)\rho(0){\cal U}^{\prime}(t)^{\dagger}
={\cal U}(t)\rho(0){\cal U}(t)^{\dagger}=\rho(t).
\end{eqnarray}
The total phase is then transformed as 
\begin{eqnarray}
\gamma_{T}\rightarrow \gamma_{T}^{\prime}={\rm arg}
\{{\rm Tr}[\rho(0)
{\cal U}^{\prime}(T)]\}={\rm arg}\{\sum_{k}\omega_{k}
\langle k|{\cal U}(T)|k\rangle e^{i\theta_{k}(T)}\}
\end{eqnarray}
while the ``dynamical phase'' is transformed as 
\begin{eqnarray}
\gamma_{D}\rightarrow \gamma^{\prime}_{D}&=&-i\int_{0}^{T}dt
{\rm Tr}[\rho(0){\cal U}^{\prime}(t)^{\dagger}\dot{{\cal U}}^{\prime}(t)]
\nonumber\\
&=&-i\int_{0}^{T}dt
{\rm Tr}[\rho(0){\cal U}(t)^{\dagger}\dot{{\cal U}}(t)]
+\sum_{k}\omega_{k}(\theta_{k}(T)-\theta_{k}(0)).
\end{eqnarray}
The simple subtraction of the "dynamical phase" from the total
phase does not give a result invariant under the above transformation for the general non-degenerate mixed state.

To alleviate this gauge non-invariance, a manifestly gauge 
invariant expression 
\begin{eqnarray}
\gamma_{G}[{\cal U}]={\rm arg}\{\sum_{k}[\omega_{k}\langle k|
{\cal U}(T)|k\rangle\exp\{i\int_{0}^{T}dt \langle k|
{\cal U}(t)^{\dagger}i\dot{{\cal U}}(t)|k\rangle\}]\}
\end{eqnarray}
has been proposed~\cite{singh}. 
When the stronger condition (2.19) is imposed, this expression 
of $\gamma_{G}[{\cal U}]$ agrees with the total phase.

We note that the expression for $\gamma_{G}[{\cal U}]$ (2.24) is 
written as
\begin{eqnarray}
\gamma_{G}[{\cal U}]={\rm arg}\{\sum_{k}\omega_{k}[
\psi_{k}^{\dagger}(0)\psi_{k}(T)\exp\{i\int_{0}^{T}dt
\psi_{k}^{\dagger}(t)i\partial_{t}\psi_{k}(t)\}]\}
\end{eqnarray}
in terms of the Schr\"{o}dinger amplitudes defined by 
$\psi_{k}(t)={\cal U}(t)|k\rangle$ which satisfy
\begin{eqnarray}
i\hbar\partial_{t}\psi_{k}(t)=\hat{H}(t)\psi_{k}(t), \ \ \ k= 1
\sim N.
\end{eqnarray}
The formula for $\gamma_{G}[{\cal U}]$, which is non-linear (to 
be precise, not bi-linear) in the Schr\"{o}dinger amplitudes, is
thus a direct generalization of (2.11) for a pure state, and it
is invariant under the transformation (2.20), namely, the 
equivalence class 
\begin{eqnarray}
\{e^{i\theta_{k}(t)}\psi_{k}(t)\}, \ \ \ k=1 \sim N.
\end{eqnarray}
Note that $\gamma_{G}[{\cal U}]$ is also written as
\begin{eqnarray}
\gamma_{G}[{\cal U}]={\rm arg}\{\sum_{k}\omega_{k}[
\psi_{k}^{\dagger}(0)\psi_{k}(T)\exp\{\frac{i}{\hbar}
\int_{0}^{T}dt\psi_{k}^{\dagger}(t)\hat{H}(t)\psi_{k}(t)\}]\}
\end{eqnarray}
and thus one implicitly assumes the equivalence class of 
Hamiltonians
\begin{eqnarray}
\{\hat{H}(t)-\hbar\partial_{t}\theta_{k}(t)\}, \ \ \ k=1 \sim N
\end{eqnarray} 
for each state $\psi_{k}(t)$ {\em separately} corresponding to 
(2.27).

The stronger condition (2.19) is written as
\begin{eqnarray}
\psi^{\dagger}_{k}(t)i\hbar\partial_{t}\psi_{k}(t)=
\psi^{\dagger}_{k}(t)\hat{H}(t)\psi_{k}(t)=0, \ \ \ k=1 \sim N.
\end{eqnarray}
The Schr\"{o}dinger equation (2.26) is, however, not maintained
under the gauge transformation (2.27) except
for the case 
\begin{eqnarray}
\partial_{t}\theta_{1}(t)= ... 
=\partial_{t}\theta_{N}(t).
\end{eqnarray}
For this special case, the modified Schr\"{o}dinger equation 
\begin{eqnarray}
i\hbar\partial_{t}\psi^{\prime}_{k}(t)=\hat{H}^{\prime}(t)
\psi^{\prime}_{k}(t),
\ \ \ k=1 \sim N
\end{eqnarray}
with $\psi^{\prime}_{k}(t)=e^{i\theta_{k}(t)}\psi_{k}(t)$
and the $k$-independent 
$\hat{H}^{\prime}(t)=\hat{H}(t)-\hbar\partial_{t}\theta_{k}(t)$
 is satisfied. All the pure states in the density matrix in 
quantum statistical mechanics, for example, are specified by 
a single Hamiltonian, and thus the expression of the partition
function ${\rm Tr}e^{-\beta\hat{H}}$ is valid. One may impose 
the same constraint on the density matrix in the present case 
also, and thus the condition (2.31); the universal Hamiltonian 
for all the pure states and the gauge invariance requirement 
(2.27) are not compatible. The interpretation of the geometric 
phase (2.25) as the holonomy associated with the gauge 
transformation (2.27), which is a direct generalization of 
(2.11), is not compatible with the universal Hamiltonian. It is 
also significant that the physical observable (2.12) which is 
expressed in terms of the total phase and visibility 
$|{\rm Tr}[\rho(0){\cal U}(T)]|$ is not gauge invariant as is 
shown in (2.22).

\subsection{A new formulation}

We start with a given hermitian Hamiltonian $\hat{H}(t)$ and 
thus given 
${\cal U}(t)
=\\ T^{\star}\exp[-\frac{i}{\hbar}\int_{0}^{t} \hat{H}(t)dt]$ 
in (2.6). We employ a diagonal form of the density matrix
\begin{eqnarray}
\rho(0)=\sum_{k}\omega_{k}|k\rangle\langle k|, \ \ \ \ \ 
\langle k|l\rangle=\delta_{k,l}
\end{eqnarray}
and observe that 
\begin{eqnarray}
\psi^{\dagger}_{k}(t)\psi_{l}(t)=\delta_{k,l}
\end{eqnarray}
for $\psi_{k}(t)={\cal U}(t)|k\rangle$ in (2.26). We then 
have the total phases for  pure states $\psi_{k}(t)$ as
\begin{eqnarray}
\phi_{k}(t)={\rm arg}\psi^{\dagger}_{k}(0)\psi_{k}(t).
\end{eqnarray}
The complete set of basis vectors $\{v_{k}(t) \}$ in Appendix A 
may then be chosen as
\begin{eqnarray}
v_{k}(t)=e^{-i\phi_{k}(t)}\psi_{k}(t), \ \ \ \ \ 
v^{\dagger}_{k}(t)v_{l}(t)=\delta_{k,l}.
\end{eqnarray}
Note that
\begin{eqnarray} 
v^{\dagger}_{k}(0)v_{k}(t)={\rm real\ and\ positive}
\end{eqnarray}
in the present definition.

In the general $\vec{x}$-dependent case, we define the
 Schr\"{o}dinger amplitudes by
\begin{eqnarray}
\psi_{k}(t,\vec{x})=\langle\vec{x}|{\cal U}(t)|k\rangle
=\int d^{3}y\langle\vec{x}|{\cal U}(t)|\vec{y}\rangle 
v_{k}(0,\vec{y})
\end{eqnarray}
and $v_{k}(t,\vec{x})=e^{-i\phi_{k}(t)}\psi_{k}(t,\vec{x})$.
We can also write the Schr\"{o}dinger amplitudes 
as~\cite{fujikawa3}
\begin{eqnarray}
\psi_{k}(\vec{x},t)
&=&v_{k}(\vec{x},t)\\
&&\times
\exp\{-\frac{i}{\hbar}\int_{0}^{t}
[\int d^{3}x v^{\dagger}_{k}(\vec{x},t)
\hat{H}(t)v_{k}(\vec{x},t)
-\langle k|i\hbar\frac{\partial}{\partial t}|k\rangle]\}\nonumber
\end{eqnarray}
with 
\begin{eqnarray}
\langle k|i\hbar\frac{\partial}{\partial t}|k\rangle\equiv
\int d^{3}x v^{\dagger}_{k}(\vec{x},t)
i\hbar\frac{\partial}{\partial t}v_{k}(\vec{x},t).
\end{eqnarray}
The Schr\"{o}dinger amplitudes in (2.39) are non-linear in the 
basis vectors, but the non-linearity of the probability amplitudes in the basis vectors is common in field theory.
The existence of a hidden local symmetry in $\psi_{k}(t,\vec{x})$
becomes clear in the second quantized formulation, where the 
field operator is expanded by using the basis vectors in (2.36)
\begin{eqnarray}
\hat{\psi}(t,\vec{x})=\sum_{k}\hat{b}_{k}(t)v_{k}(t,\vec{x})
\end{eqnarray}
and the effective Hamiltonian in (A.4) is given by 
\begin{eqnarray}
\hat{H}_{eff}(t)&=&\sum_{n}\hat{b}_{n}^{\dagger}(t)[
\int d^{3}x v^{\dagger}_{n}(\vec{x},t)
\hat{H}(t)v_{n}(\vec{x},t)
\nonumber\\
&&-
\langle n|i\hbar\frac{\partial}{\partial t}|n\rangle]
\hat{b}_{n}(t)
\end{eqnarray}
which is diagonal. The operator $\hat{\psi}(t,\vec{x})$ is 
invariant under the hidden local gauge symmetry
\begin{eqnarray}
v_{k}(\vec{x},t)\rightarrow e^{i\alpha_{k}(t)}
v_{k}(\vec{x},t), \ \ \ \ \ \hat{b}_{k}(t)\rightarrow 
e^{-i\alpha_{k}(t)}\hat{b}_{k}(t)
\end{eqnarray}
with a general function $\alpha_{k}(t)$. Any formulation 
should preserve this exact symmetry. The Schr\"{o}dinger
amplitude $\psi_{k}(t,\vec{x})$ in (2.39) is transformed under 
the hidden local symmetry as 
\begin{eqnarray}
\psi_{k}(t,\vec{x})\rightarrow e^{i\alpha_{k}(0)}
\psi_{k}(t,\vec{x})
\end{eqnarray}
independently of $t$ and thus the Schr\"{o}dinger equation with 
a fixed Hamiltonian is invariant under the hidden local symmetry.
This property is clearly seen in the general expression (A.6)
in Appendix. The difference between the equivalence class (2.27)
and the present hidden local gauge symmetry is obvious. From the
 point of view of the hidden local symmetry, the condition 
(2.37) is realized by a specific choice of the hidden local 
gauge.

The quantity ${\rm Tr}{\cal U}(T)\rho(0)$ appearing in (2.12) is
 then  written as 
\begin{eqnarray}
{\rm Tr}{\cal U}(T)\rho(0)&=&\sum_{k}\omega_{k}
\psi_{k}^{\dagger}(0,\vec{x})\psi_{k}(T,\vec{x})\nonumber\\
&=&\sum_{k}\omega_{k}
v_{k}^{\dagger}(0,\vec{x})v_{k}(T,\vec{x})
\exp\{\frac{i}{\hbar}\int_{0}^{T}dtd^{3}x[
v_{k}^{\dagger}(t,\vec{x})i\hbar\partial_{t}v_{k}(t,\vec{x})
\nonumber\\
&&-v_{k}^{\dagger}(t,\vec{x})\hat{H}(t)v_{k}(t,\vec{x})]\}
\end{eqnarray}
by using the density matrix 
$\rho(0)=\sum_{k}\omega_{k}
\psi_{k}(0,\vec{x})\psi^{\dagger}_{k}(0,\vec{x})$. This quantity
is manifestly invariant under the hidden local symmetry with a 
fixed Hamiltonian, and thus not only the total phase (2.15) but
also the visibility $|{\rm Tr}{\cal U}(T)\rho(0)|$ in (2.12) are
 invariant. This expression of ${\rm Tr}{\cal U}(T)\rho(0)$ 
contains the holonomy
\begin{eqnarray}
\bar{v}^{\dagger}_{k}(0,\vec{x})\bar{v}_{k}(T,\vec{x})
\end{eqnarray}
for
\begin{eqnarray}
\bar{v}_{k}(t,\vec{x})=v_{k}(t,\vec{x})
\exp\{\frac{i}{\hbar}\int_{0}^{t}dt
\langle k|i\hbar\partial_{t}|k\rangle]\}.
\end{eqnarray}
This holonomy is fixed by the hidden local symmetry. If one 
considers $\bar{v}_{k}(t,\vec{x})
=e^{i\alpha_{k}(t)}v_{k}(t,\vec{x})$ by using the hidden local
symmetry, the parallel transport condition
\begin{eqnarray}
\int d^{3}x\bar{v}^{\dagger}_{k}(t,\vec{x})
i\hbar\partial_{t}\bar{v}_{k}(t,\vec{x})=0
\end{eqnarray} 
fixes $\alpha_{k}(t)$ as in (2.47).

If all the pure states perform cyclic evolution with the same 
period $T$, one can choose the hidden local gauge such that
\begin{eqnarray}
v_{k}^{\dagger}(0,\vec{x})v_{k}(T,\vec{x})={\rm real\ and\
positive}
\end{eqnarray}
for all $k$, and the exponential factor in (2.45) exhibits 
the entire geometrical phase together with the ``dynamical phase'' $(1/\hbar)\int_{0}^{T}dtd^{3}xv_{k}^{\dagger}(t,\vec{x})
\hat{H}(t)v_{k}(t,\vec{x})$ of each 
pure state. In the case of the cyclic
evolution,  one can use the formula (2.45) for the spatial 
interference pattern analogous to the Aharonov-Bohm phase.
In practice, the cyclic evolution of all the pure states 
$\psi_{k}(t)$ with a period $T$ may be rather exceptional. For a
generic case, we need to define the total phase for  
non-cyclic evolution~\cite{samuel} as the phase of
\begin{eqnarray}
{\rm Tr}{\cal U}(T)\rho(0)&
=&\sum_{k}\omega_{k}\int d^{3}x
\psi_{k}^{\dagger}(0,\vec{x})\psi_{k}(T,\vec{x})\\
&=&\sum_{k}\omega_{k}\int d^{3}x
v_{k}^{\dagger}(0,\vec{x})v_{k}(T,\vec{x})\nonumber\\
&&\times\exp\{\frac{i}{\hbar}\int_{0}^{T}dtd^{3}x[
v_{k}^{\dagger}(t,\vec{x})i\hbar\partial_{t}v_{k}(t,\vec{x})
-v_{k}^{\dagger}(t,\vec{x})\hat{H}(t)v_{k}(t,\vec{x})]\}
\nonumber
\end{eqnarray}
where we included the integration over the spatial coordinates
$\vec{x}$ in the trace by following the analysis 
in~\cite{fujikawa}. Note that (2.37), which is realized by a 
suitable choice of 
the hidden local gauge, is written in the present notation 
as $\int d^{3}x v_{k}^{\dagger}(0,\vec{x})v_{k}(t,\vec{x})=
{\rm real\ and\ positive}$.

A salient feature of the present formulation is that the total
phase and visibility in (2.45) or (2.50), which are the direct
 observables in the 
interference experiment~\cite{sjoqvist}, are gauge invariant, 
and the geometric phase arises from the holonomy of the basis 
vectors rather than from the Schr\"{o}dinger amplitudes. The 
Schr\"{o}dinger equation is invariant under the hidden local
symmetry with a fixed Hamiltonian. The parallel transport 
condition on the Schr\"{o}dinger amplitudes constrains the 
Hamiltonian as in (2.30), whereas the parallel transport of the 
basis vectors is realized  without any constraint on the 
Hamiltonian as in (2.47). In the present formulation, the 
geometric phase is analyzed on the basis of a given general 
Hamiltonian instead of imposing constraints on the Hamiltonian, 
which is in accord with the original idea of the non-adiabatic 
phase~\cite{aharonov, samuel}. For some applications in quantum
information~\cite{jones,falci,fuentes,ekert,pachos}, 
for example, one may 
be interested in the inverse, namely, in finding a Hamiltonian 
which reproduces a given density matrix~\cite{tong}. In such a 
case, one may impose the 
condition (2.30) also. In any case, if the stronger condition 
(2.30) happens to be satisfied for a specific Hamiltonian,
our formula for the total phase gives the geometric phase which 
is the same as in the past formulation.

\section{Example: Spin polarization}

We analyze a simple example described by a time dependent 
Hamiltonian 
\begin{eqnarray}
&&\hat{H}=-\mu\hbar\vec{B}(t)\vec{\sigma},\nonumber\\
&&\vec{B}(t)=B(\sin\theta\cos\varphi(t), 
\sin\theta\sin\varphi(t),\cos\theta )
\end{eqnarray}
where $\varphi(t)=\omega t$ with constant $\omega$, $B$ and 
$\theta$.
The effective Hamiltonian in (A.4) then becomes 
\begin{eqnarray}
\hat{H}_{eff}(t)&=&[-\mu\hbar B
-\frac{(1+\cos\theta)}{2}\hbar\omega]\hat{b}^{\dagger}_{+}
\hat{b}_{+}
+[\mu\hbar B-\frac{1-\cos\theta}{2}\hbar\omega]
\hat{b}^{\dagger}_{-}\hat{b}_{-}
\nonumber\\
&-&\frac{\sin\theta}{2}\hbar\omega
[\hat{b}^{\dagger}_{+}\hat{b}_{-}+
\hat{b}^{\dagger}_{-}\hat{b}_{+}]
\end{eqnarray}
if one uses the basis set 
\begin{eqnarray}
v_{+}(t)=\left(\begin{array}{c}
            \cos\frac{1}{2}\theta e^{-i\varphi(t)}\\
            \sin\frac{1}{2}\theta
            \end{array}\right), \ \ \ \ \ 
v_{-}(t)=\left(\begin{array}{c}
            \sin\frac{1}{2}\theta e^{-i\varphi(t)}\\
            -\cos\frac{1}{2}\theta
            \end{array}\right)
\end{eqnarray}
which satisfy $\hat{H}(t)v_{\pm}(t)=\mp\mu\hbar Bv_{\pm}(t)$
and the relations
\begin{eqnarray}
v^{\dagger}_{+}(t)i\frac{\partial}{\partial t}v_{+}(t)
&=&\frac{(1+\cos\theta)}{2}\omega
\nonumber\\
v^{\dagger}_{+}(t)i\frac{\partial}{\partial t}v_{-}(t)
&=&\frac{\sin\theta}{2}\omega
=v^{\dagger}_{-}(t)i\frac{\partial}{\partial t}v_{+}(t)
,\nonumber\\
v^{\dagger}_{-}(t)i\frac{\partial}{\partial t}v_{-}(t)
&=&\frac{1-\cos\theta}{2}\omega.
\end{eqnarray}
The above effective Hamiltonian (3.2) is not diagonal, and thus
 quantum state mixing takes place if one uses the basis set
$v_{\pm}(t)$. 
We thus perform a unitary transformation
\begin{eqnarray}
\left(\begin{array}{c}
     \hat{b}_{+}(t)\\
     \hat{b}_{-}(t)
     \end{array}\right)
&=&
\left(\begin{array}{cc}
 \cos\frac{1}{2}\alpha&-\sin\frac{1}{2}\alpha\\
 \sin\frac{1}{2}\alpha &\cos\frac{1}{2}\alpha
            \end{array}\right)
\left(\begin{array}{c}
           \hat{c}_{+}(t)\\
           \hat{c}_{-}(t)
            \end{array}\right)\nonumber\\
&\equiv&U^{T}\left(\begin{array}{c}
           \hat{c}_{+}(t)\\
           \hat{c}_{-}(t)
            \end{array}\right)
\end{eqnarray}
where $U^{T}$ stands for the transpose of $U$. 
The eigenfunctions are transformed to
\begin{eqnarray}
\left(\begin{array}{c}
     w_{+}(t)\\
     w_{-}(t)
     \end{array}\right)
&=&U\left(\begin{array}{c}
           v_{+}(t)\\
           v_{-}(t)
            \end{array}\right)\nonumber\\
&=&
\left(\begin{array}{cc}
 \cos\frac{1}{2}\alpha&\sin\frac{1}{2}\alpha\\
 -\sin\frac{1}{2}\alpha &\cos\frac{1}{2}\alpha
            \end{array}\right)
\left(\begin{array}{c}
           v_{+}(t)\\
           v_{-}(t)
            \end{array}\right)
\end{eqnarray}
or explicitly
\begin{eqnarray}
w_{+}(t)=\left(\begin{array}{c}
            \cos\frac{1}{2}(\theta-\alpha) e^{-i\varphi(t)}\\
            \sin\frac{1}{2}(\theta-\alpha)
            \end{array}\right), \ \ \ \ \ 
w_{-}(t)=\left(\begin{array}{c}
            \sin\frac{1}{2}(\theta-\alpha) e^{-i\varphi(t)}\\
            -\cos\frac{1}{2}(\theta-\alpha)
            \end{array}\right).
\end{eqnarray}
The field variable $\hat{\psi}(t,\vec{x})$ in second quantization
is given by 
\begin{eqnarray}
\hat{\psi}(t,\vec{x})&=&\sum_{n=\pm}\hat{b}_{n}(t)v_{n}(t)
\nonumber\\
&=&\sum_{n=\pm}\hat{c}_{n}(t)w_{n}(t)
\end{eqnarray}
which is invariant under the hidden local symmetry
\begin{eqnarray}
w_{n}(t)\rightarrow e^{i\alpha_{n}(t)}w_{n}(t), \ \ \ \
\hat{c}_{n}(t)\rightarrow e^{-i\alpha_{n}(t)}\hat{c}_{n}(t).
\end{eqnarray}
We also have
\begin{eqnarray}
&&w_{\pm}^{\dagger}(t)\hat{H}w_{\pm}(t)
=\mp \mu\hbar H\cos\alpha\nonumber\\
&&w_{\pm}^{\dagger}(t)i\hbar\partial_{t}w_{\pm}(t)
=\frac{\hbar\omega}{2}(1\pm\cos(\theta-\alpha)).
\end{eqnarray}

If one chooses the parameter $\alpha$ in (3.5) as 
\begin{eqnarray}
\tan\alpha=\frac{\hbar\omega\sin\theta}{2\mu\hbar B+\hbar\omega
\cos\theta}
\end{eqnarray}
or equivalently 
$2\mu\hbar B\sin\alpha=\hbar\omega\sin(\theta-\alpha)$, one 
obtains a diagonal effective Hamiltonian~\cite{fujikawa3} 
\begin{eqnarray}
\hat{H}_{eff}(t)&=&\sum_{n=\pm}\hat{c}^{\dagger}_{n}
[\mp\mu\hbar B\cos\alpha
-\frac{\hbar\omega}{2}(1\pm\cos(\theta-\alpha))]\hat{c}_{n}
\nonumber\\
&=&\sum_{n=\pm}\hat{c}^{\dagger}_{n}
[w_{n}^{\dagger}(t)\hat{H}w_{n}(t)
-w_{n}^{\dagger}(t)i\hbar\partial_{t}w_{n}(t)]\hat{c}_{n}
\end{eqnarray}
and thus the quantum state mixing is avoided. 
The above unitary transformation is time-independent and thus 
the effective Hamiltonian is not changed 
$\hat{H}_{eff}(b^{\dagger}_{\pm}(t),b_{\pm}(t))
=\hat{H}_{eff}(c^{\dagger}_{\pm}(t),c_{\pm}(t))$.
We  have the Schr\"{o}dinger amplitudes with initial
conditions $\psi_{\pm}(0)=
w_{\pm}(0)$ as (see (A.7) in Appendix)
\begin{eqnarray}
\psi_{\pm}(t)
&=&w_{\pm}(t)\exp\{-\frac{i}{\hbar}[\mp\mu\hbar B\cos\alpha
-\frac{\hbar\omega}{2}(1\pm\cos(\theta-\alpha))]t\}
\nonumber\\
&=&w_{\pm}(t)\exp\{-\frac{i}{\hbar}\int_{0}^{t}dt
[w_{\pm}^{\dagger}(t)\hat{H}w_{\pm}(t)
-w_{\pm}^{\dagger}(t)i\hbar\partial_{t}w_{\pm}(t)]\}.
\end{eqnarray}
These expressions are periodic with a period  
$T=\frac{2\pi}{\omega}$ up to a phase by noting 
$w_{\pm}(0)=w_{\pm}(T)$, and they are exact. 
From the point of view of the diagonalization 
of the Hamiltonian, we have not completely
diagonalized the Hamiltonian since $w_{\pm}(t)$ carry
certain time-dependence. These formulas $\psi_{\pm}(t)$ are 
invariant under the hidden local symmetry (3.9) up to a constant
 phase factor, 
$\psi_{\pm}(t)\rightarrow e^{i\alpha_{\pm}(0)}\psi_{\pm}(t)$, 
and the geometric phase is identified by means of the analysis 
of hidden local symmetry as the phase of 
\begin{eqnarray}
\exp\{-\frac{i}{\hbar}\int_{0}^{T}dt
[-w^{\dagger}_{\pm}(t)i\hbar\partial_{t}w_{\pm}(t)]\}
=\exp\{-i\pi(1\mp\cos(\theta-\alpha)) \}
\end{eqnarray}
up to $2n\pi$.

One can measure $\psi^{\dagger}_{+}(0)\psi_{+}(T)$, for example,
 which is manifestly invariant under the hidden local symmetry 
by the interference in
\begin{eqnarray}
\frac{1}{2}|\psi_{+}(T)+\psi_{+}(0)|^{2}
&=&|\psi_{+}(0)|^{2}+{\rm Re}\psi^{\dagger}_{+}(0)\psi_{+}(T)
\nonumber\\
&=&1+\cos[(\mu B
\cos\alpha)T
-\frac{1}{2}\Omega_{+}]
\end{eqnarray}
where
\begin{eqnarray}
\Omega_{+}=2\pi [1-\cos(\theta-\alpha)]
\end{eqnarray}
stands for the solid angle subtended by 
$w_{+}^{\dagger}(t)\vec{\sigma}w_{+}(t)$ by noting
\begin{eqnarray}
w_{+}^{\dagger}(t)\vec{\sigma}w_{+}(t)
&=&(\sin(\theta-\alpha)\cos\varphi,\sin(\theta-\alpha)
\sin\varphi,\cos(\theta-\alpha))
\nonumber\\
&=&-w_{-}^{\dagger}(t)\vec{\sigma}w_{-}(t).
\end{eqnarray}
The separation of the non-adiabatic phase 
$\frac{1}{2}\Omega_{+}$ and the ``dynamical phase'' 
$(\mu B\cos\alpha)T$ in (3.15) is achieved by varying the 
parameters in the Hamiltonian, namely, $B$, $\omega$ and 
$\theta$. In passing, we 
note that the non-adiabatic phase $\frac{1}{2}\Omega_{+}$ in 
(3.16) (or non-adiabatic phase in general) is 
topologically trivial~\cite{fujikawa3} and thus the stability 
argument on the basis of topology is not used.

To define the basis set to represent any state $|k\rangle$ of 
the initial density matrix $\rho(0)=\sum |k\rangle\langle k|$ in
 (2.33), one may define linear combinations
\begin{eqnarray}
\Psi_{+}(t)&=&\cos\frac{\Theta}{2} \psi_{+}(t)
+\sin\frac{\Theta}{2} \psi_{-}(t)\nonumber\\
&=&\tilde{w}_{+}(t)
\exp\{-\frac{i}{\hbar}[-\mu\hbar B\cos\alpha
-\frac{\hbar\omega}{2}(1+\cos(\theta-\alpha))]t\}
\nonumber\\
&=&\tilde{w}_{+}(t)\exp\{-\frac{i}{\hbar}\int_{0}^{t}dt
[\tilde{w}_{+}^{\dagger}(t)\hat{H}\tilde{w}_{+}(t)
-\tilde{w}_{+}^{\dagger}(t)i\hbar\partial_{t}\tilde{w}_{+}(t)]\},
\nonumber\\
\Psi_{-}(t)&=&-\sin\frac{\Theta}{2} \psi_{+}(t)
+\cos\frac{\Theta}{2} \psi_{-}(t)\nonumber\\
&=&\tilde{w}_{-}(t)
\exp\{-\frac{i}{\hbar}[\mu\hbar B\cos\alpha
-\frac{\hbar\omega}{2}(1-\cos(\theta-\alpha))]t\}
\nonumber\\
&=&\tilde{w}_{-}(t)\exp\{-\frac{i}{\hbar}\int_{0}^{t}dt
[\tilde{w}_{-}^{\dagger}(t)\hat{H}\tilde{w}_{-}(t)
-\tilde{w}_{-}^{\dagger}(t)i\hbar\partial_{t}
\tilde{w}_{-}(t)]\}
\end{eqnarray}
where
\begin{eqnarray}
\tilde{w}_{+}(t)&=&\cos\frac{\Theta}{2} w_{+}(t)+
\sin\frac{\Theta}{2} w_{-}(t)\exp\{-\frac{i}{\hbar}
[2\mu\hbar B\cos\alpha
+\hbar\omega(\cos(\theta-\alpha))]t\},\nonumber\\
\tilde{w}_{-}(t)&=&-\sin\frac{\Theta}{2} w_{+}(t)
\exp\{\frac{i}{\hbar}[2\mu\hbar B\cos\alpha
+\hbar\omega(\cos(\theta-\alpha))]t\}+
\cos\frac{\Theta}{2} w_{-}(t).
\end{eqnarray}
These $\Psi_{\pm}(t)$ satisfy the Schr\"{o}dinger equation, and 
one can represent any initial state $|k\rangle$ by choosing 
$\Theta$ suitably. The basis vectors $\tilde{w}_{\pm}(t)$ satisfy
$\tilde{w}^{\dagger}_{n}(t)\tilde{w}_{m}(t)=\delta_{n,m}$ and 
\begin{eqnarray}
\tilde{w}_{\pm}^{\dagger}(t)\hat{H}\tilde{w}_{\pm}(t)
&=&\mp \mu\hbar B[\cos\alpha\cos\Theta-\sin\alpha\sin\Theta\cos\beta(t)], \nonumber\\
\tilde{w}_{\pm}^{\dagger}(t)i\hbar\partial_{t}
\tilde{w}_{\pm}(t)
&=&\pm \mu\hbar B[\cos\alpha(1-\cos\Theta)+\sin\alpha\sin\Theta\cos\beta(t)]
\nonumber\\
&&+\frac{\hbar\omega}{2}(1\pm\cos(\theta-\alpha))
\end{eqnarray}
where $\beta(t)=[2\mu B\cos\alpha
+\omega\cos(\theta-\alpha)]t$.

The basis set $\tilde{w}_{\pm}(t)$ are not periodic in general, 
and they are periodic with a period $T$ only 
when 
\begin{eqnarray}
&&T\omega=2\pi n,\nonumber\\
&&T[2\mu B\cos\alpha
+\omega\cos(\theta-\alpha)]=2\pi m
\end{eqnarray}
with two integers $n$ and $m$. Related to this property, the 
geometric phase for a pure state $\Psi_{+}(t)$, for example,
\begin{eqnarray}
\Psi^{\dagger}_{+}(0)\Psi_{+}(T)
&=&\tilde{w}^{\dagger}_{+}(0)\tilde{w}_{+}(T)\nonumber\\
&\times&\exp\{-\frac{i}{\hbar}\int_{0}^{T}dt
[\tilde{w}_{+}^{\dagger}(t)\hat{H}\tilde{w}_{+}(t)
-\tilde{w}_{+}^{\dagger}(t)i\hbar\partial_{t}\tilde{w}_{+}(t)]\}
\end{eqnarray}
arises not only from the phase factor, {\it i.e.}, the second 
term on the exponential but also
from the pre-factor $\tilde{w}^{\dagger}_{+}(0)\tilde{w}_{+}(T)$
for a general $T$ which does not satisfy (3.21). One may utilize 
the hidden local gauge symmetry 
$\tilde{w}_{+}(t)\rightarrow e^{i\tilde{\alpha}_{+}(t)}
\tilde{w}_{+}(t)$, under which the quantity (3.22) is invariant, 
to make the pre-factor 
$\tilde{w}^{\dagger}_{+}(0)\tilde{w}_{+}(T)$ real and positive,
and then only the  phase factor on the exponential contributes
to the geometric phase.

\subsection{Mixed states}

We study a specific density matrix defined by a pure state in 
(3.18)
\begin{eqnarray}
\rho(t)=|\Psi_{+}(t)\rangle\langle\Psi_{+}(t)|
\end{eqnarray}
which is not cyclic in general. The density matrix for a mixed
state may be defined by 
\begin{eqnarray}
\rho_{mix}(t)=\cos^{2}\frac{\Theta}{2}|\psi_{+}(t)\rangle
\langle\psi_{+}(t)|+\sin^{2}\frac{\Theta}{2}|\psi_{-}(t)\rangle
\langle\psi_{-}(t)|
\end{eqnarray}
by assuming the de-coherence of $\psi_{\pm}(t)$, and 
$\rho_{mix}(t)$ is cyclic with a period $T$.

The total phase may be defined as the phase of~\cite{sjoqvist}
\begin{eqnarray}
&&{\rm Tr}\{T^{\star}e^{-\frac{i}{\hbar}\int_{0}^{T}dt\hat{H}(t)}
\rho_{mix}(0)\}\nonumber\\
&&=\cos^{2}\frac{\Theta}{2}\langle\psi_{+}(0)|\psi_{+}(T)\rangle
+\sin^{2}\frac{\Theta}{2}\langle\psi_{-}(0)|\psi_{-}(T)\rangle
\\
&&=\cos^{2}\frac{\Theta}{2}w^{\dagger}_{+}(0)w_{+}(T)
\exp\{-\frac{i}{\hbar}\int_{0}^{T}dt
[w_{+}^{\dagger}(t)\hat{H}w_{+}(t)
-w_{+}^{\dagger}(t)i\hbar\partial_{t}w_{+}(t)]\}\nonumber\\
&&+\sin^{2}\frac{\Theta}{2}w^{\dagger}_{-}(0)w_{-}(T)
\exp\{-\frac{i}{\hbar}\int_{0}^{T}dt
[w_{-}^{\dagger}(t)\hat{H}w_{-}(t)
-w_{-}^{\dagger}(t)i\hbar\partial_{t}w_{-}(t)]\}\nonumber
\end{eqnarray}
which is manifestly invariant under the hidden local symmetry
$w_{\pm}(t)\rightarrow e^{i\alpha_{\pm}(t)}w_{\pm}(t)$ and 
one can uniquely identify the geometric phase for each pure 
state $\psi_{\pm}(t)$. In the present case, one can in principle
 distinguish the geometric phase from the ``dynamical phase'' 
contained in the total phase by varying the 
parameters $T=2\pi/\omega$, $B$, $\theta$ and the angle 
$\Theta$. See also~\cite{falci, fuentes, ekert}.  
Note that we can in principle separate the geometric phase and 
the ``dynamical phase'' by varying the parameters 
$T=2\pi/\omega$, $B$ and $\theta$ for pure states 
$\psi_{\pm}(t)$ as in (3.15). 

As a special example, one may choose $\alpha=\pi/2$ in (3.11), 
namely
\begin{eqnarray}
2\mu\hbar B+\hbar\omega\cos\theta=0.
\end{eqnarray}
Then $w_{\pm}^{\dagger}(t)\hat{H}(t)w_{\pm}(t)=0$ in 
(3.10), and only the geometric phases remain in (3.25),
\begin{eqnarray}
{\rm Tr}\{T^{\star}e^{-\frac{i}{\hbar}\int_{0}^{T}dt\hat{H}(t)}
\rho_{mix}(0)\}
&=&\cos^{2}\frac{\Theta}{2}
e^{-i\pi(1+\cos(\theta-\alpha))}\nonumber\\
&+&\sin^{2}\frac{\Theta}{2}
e^{-i\pi(1-\cos(\theta-\alpha))}
\end{eqnarray}
by using (3.14) and the periodicity $w_{\pm}(T)=w_{\pm}(0)$.
One also has
\begin{eqnarray}
\psi_{\pm}^{\dagger}(t)\hat{H}(t)\psi_{\pm}(t)=0
\end{eqnarray}
and thus the parallel transport condition (2.30) in the 
conventional formulation is realized by a suitable choice of 
the Hamiltonian instead of the gauge transformation (2.27). 
This choice of parameters corresponds to an explicit example 
in~\cite{tong}.

In the example discussed here, the explicit form of basis 
vectors $w_{\pm}(t)$ in (3.7) was defined by a diagonalization 
of the effective Hamiltonian which is a natural generalization 
of the analysis of adiabatic phase, instead of the construction 
discussed in Section 2.2. The periodicity condition 
$w_{\pm}(T)=w_{\pm}(0)$ with  $T=2\pi/\omega$ severely 
constrains the possible initial state $|k\rangle$ of 
$\rho(0)=\sum_{k}\omega_{k}|k\rangle\langle k|$ in (2.33).
For generic states $|k\rangle$ one may use the basis vectors
$\tilde{w}_{\pm}(t)$ in (3.19). The basis vectors 
$\tilde{w}_{\pm}(t)$ generally agree with the basis vectors 
$v_{k}(t)$ in (2.36) up to hidden local gauge symmetry.
 
\section{Discussion}

The mixed state generally appears as a result of de-coherence in 
the actual experimental situation. A precise description of 
de-coherent processes in quantum mechanics is involved. 
Operationally, one may reduce a pure state to a mixed state by
 random phase approximation, for example. A more elegant 
procedure is the purification and its inverse, namely, 
reduction. See, for example, Ref.~\cite{uhlmann}. The basic idea 
of the reduction starts with a normalized pure (entangled) state
\begin{eqnarray}
\psi(t)=\sum_{n,m}a_{n,m}|\psi_{n}(t)\rangle|\phi_{m}(t)\rangle
\end{eqnarray}
and a density matrix for the pure state
\begin{eqnarray}
\rho(t)=|\psi(t)\rangle\langle\psi(t)|.
\end{eqnarray}
One then takes a partial trace over the states 
$|\phi_{m}(t)\rangle$ and obtains a density matrix for a mixed 
state
\begin{eqnarray}
\rho_{mix}(t)&=&{\rm Tr}_{\phi}\{\rho(t)\}\nonumber\\
&=&\sum_{n,l}\omega_{n,l}|\psi_{n}(t)\rangle
\langle\psi_{l}(t)|
\end{eqnarray}
where
\begin{eqnarray}
\omega_{n,l}\equiv \sum_{m}a_{n,m}a^{\star}_{l.m}.
\end{eqnarray}
If the dimension of the states $\{|\psi_{n}(t)\rangle\}$ is $N$, 
one may choose the dimension of the states 
$\{|\phi_{m}(t)\rangle\}$ to 
be equal to or larger than $N$. One then has more than $N^{2}$
free parameters $a_{n,m}$ which are naively sufficient to 
describe $N^{2}$ parameters $\omega_{n,l}$. The purification is 
the inverse of the above procedure and corresponds to a 
construction of a pure state starting with a mixed state, though
the purification is not unique.   

The time development of the pure state is given by
\begin{eqnarray}
i\hbar\partial_{t}\psi(t)=[\hat{H}+\hat{\tilde{H}}]\psi(t)
\end{eqnarray}
where $\hat{H}$ acts on the states $\{|\psi_{n}(t)\rangle\}$ and 
$\hat{\tilde{H}}$ on the states $\{|\phi_{m}(t)\rangle\}$. 
The pure state (4.1) is transformed under the equivalence class 
$\{e^{i\theta_{k}(t)}\psi_{k}(t)\}$ in (2.27) as
\begin{eqnarray}
\psi(t) \rightarrow \psi^{\prime}(t)
=\sum_{n,m}a_{n,m}e^{i\theta_{n}(t)}
|\psi_{n}(t)\rangle|\phi_{m}(t)\rangle
\end{eqnarray}
and thus the Schr\"{o}dinger equation is not maintained.
To satisfy the Schr\"{o}dinger equation one needs to consider 
an equivalence class of Hamiltonians
\begin{eqnarray}
\{\hat{H}-\hbar\partial_{t}\theta_{k}(t)\}
\end{eqnarray}
for {\em each} state $|\psi_{n}(t)\rangle$ separately. To maintain the
universal Hamiltonian for all the states contained in (4.5) only
 a limited set of gauge 
transformations (2.31) is allowed. As a consequence, one cannot 
achieve the parallel transport for all the states (2.30)
by means of the gauge transformation, which implies that the 
geometric phase (2.25) is not interpreted as the holonomy 
associated with (2.27) if one wants to maintain the 
universal Hamiltonian. The limited set of gauge 
transformations (2.31) can achieve only (2.17).
(This property is related 
to the general incompatibility of the equivalence class (2.27)
with the superposition principle such as (4.1), as analyzed 
in~\cite{fujikawa3}.) 
When one adopts the 
diagonal form of the density matrix (2.14) it is manifestly 
invariant under the equivalence class as in (2.21), but one 
needs to use a 
transformed Hamiltonian to define the physical total phase in 
(2.22). The invariance of the density matrix does not necessarily imply the gauge invariance of physical observables.  

The hidden local gauge symmetry discussed in the present
paper transforms the state vectors as 
\begin{eqnarray}
\{|\psi_{n}(t)\rangle\} \rightarrow 
\{e^{i\alpha_{n}(0)}|\psi_{n}(t)\rangle\} 
\end{eqnarray}
and thus the pure state (4.1)
\begin{eqnarray}
\psi(t) \rightarrow \psi^{\prime}(t)
=\sum_{n,m}a_{n,m}e^{i\alpha_{n}(0)}
|\psi_{n}(t)\rangle|\phi_{m}(t)\rangle.
\end{eqnarray} 
The hidden local symmetry thus maps one allowed solution of 
the Schr\"{o}dinger equation to another allowed solution with
a fixed Hamiltonian, 
although the  pure state itself is changed. But the 
diagonal density matrix and the physically observable 
interference in the present application are kept invariant under
 this transformation as in (2.45) or (2.50).

In conclusion, we have illustrated the advantage of the use of 
the hidden local gauge symmetry, which is clearly recognized in 
the second quantization,  in the analyses of geometric phases
both for pure and mixed states.
As for other applications of the second quantized formulation, 
it has been shown elsewhere~\cite{fujikawa4} that the geometric 
phase and the quantum anomaly, which have been long considered 
to be closely related, have in fact little to do with  
each other.

\appendix

\section{Hidden local gauge symmetry}

In this appendix, for the sake of completeness, we recapitulate the basic idea of hidden local gauge
symmetry. We start with the generic hermitian Hamiltonian 
$\hat{H}=\hat{H}(\hat{\vec{p}},\hat{\vec{x}},X(t))$
for a single particle theory in the  background 
variable $X(t)=(X_{1}(t),X_{2}(t),...)$.
We then define a complete set of (time-dependent) eigenfunctions
\begin{eqnarray}
\int d^{3}x v^{\dagger}_{n}(\vec{x},t)v_{m}(\vec{x},t)
=\delta_{n,m},
\end{eqnarray}
which are arbitrary at this moment. 
We take the time $T$ as a period of the variable $X(t)$ and the 
basis set $\{v_{n}(t,\vec{x}) \}$ in the 
analysis of geometric phases, unless stated otherwise. 

We then expand the field 
variable $\hat{\psi}(t,\vec{x})$ in the 
notation of second quantization as
\begin{eqnarray}
\hat{\psi}(t,\vec{x})=\sum_{n}\hat{b}_{n}(t)v_{n}(\vec{x},t).
\end{eqnarray}
By using the above expansion in the action 
\begin{eqnarray}
S=\int dtd^{3}x\{\hat{\psi}^{\dagger}i\hbar\partial_{t}\hat{\psi}
-\hat{\psi}^{\dagger}\hat{H}\hat{\psi} \}
\end{eqnarray}
we obtain the effective Hamiltonian (depending on Bose or 
Fermi statistics)
\begin{eqnarray}
\hat{H}_{eff}(t)&=&\sum_{n,m}\hat{b}_{n}^{\dagger}(t)[
\int d^{3}x v^{\dagger}_{n}(\vec{x},t)
\hat{H}v_{m}(\vec{x},t)
\nonumber\\
&&-
\langle n|i\hbar\frac{\partial}{\partial t}|m\rangle]
\hat{b}_{m}(t)
\end{eqnarray}
with $[\hat{b}_{n}(t), \hat{b}^{\dagger}_{m}(t)]_{\mp}
=\delta_{n,m}$. 
The second term in the effective Hamiltonian is defined by 
\begin{eqnarray}
\int d^{3}x v^{\dagger}_{n}(\vec{x},t)
i\hbar\frac{\partial}{\partial t}v_{m}(\vec{x},t)
\equiv \langle n|i\hbar\frac{\partial}{\partial t}|m\rangle.
\end{eqnarray}
 
The probability amplitude which satisfies the Schr\"{o}dinger 
equation with $\psi_{n}(\vec{x},0)=v_{n}(\vec{x};0)$
is given by
\begin{eqnarray}
\psi_{n}(\vec{x},t)=
\langle 0|\hat{\psi}(t,\vec{x})\hat{b}^{\dagger}_{n}(0)|0\rangle
\end{eqnarray}
since $i\hbar\partial_{t}\hat{\psi}=\hat{H}\hat{\psi}$ in the 
present problem.
When one defines the Schr\"{o}dinger picture 
$\hat{{\cal H}}_{eff}(t)$ by replacing all $\hat{b}_{n}(t)$ by
$\hat{b}_{n}(0)$ in the above $\hat{H}_{eff}(t)$, one can 
write~\cite{fujikawa, fujikawa2}
\begin{eqnarray}
\psi_{n}(\vec{x},t)
&=&\sum_{m} v_{m}(\vec{x},t)\nonumber\\
&&\times
\langle m|T^{\star}\exp\{-\frac{i}{\hbar}\int_{0}^{T}
\hat{{\cal H}}_{eff}(t)
dt\}|n\rangle
\end{eqnarray}
where $T^{\star}$ stands for the time ordering operation, and
the state vectors in the second quantization are defined by $
|n\rangle=\hat{b}_{n}^{\dagger}(0)|0\rangle$. This formula is 
exact.

Our formulation contains an exact hidden local gauge symmetry
which keeps the field variable $\hat{\psi}(t,\vec{x})$ invariant 
\begin{eqnarray}
&&v_{n}(\vec{x},t)\rightarrow v^{\prime}_{n}(\vec{x},t)=
e^{i\alpha_{n}(t)}v_{n}(\vec{x},t),\nonumber\\
&&\hat{b}_{n}(t) \rightarrow \hat{b}^{\prime}_{n}(t)=
e^{-i\alpha_{n}(t)}\hat{b}_{n}(t), \ \ \ \ n=1,2,3,...,
\end{eqnarray}
where the gauge parameter $\alpha_{n}(t)$ is a general 
function of $t$. 
This gauge symmetry (or substitution rule) states the fact 
that the choice of coordinates in the functional space is 
arbitrary and this symmetry by itself does not give any 
conservation law. This symmetry is exact under a 
rather mild condition that the basis set (A.1) is not singular, 
and consequently the physical observables 
should always respect this symmetry.

Our next observation is that $\psi_{n}(\vec{x},t)$ is
transformed under the hidden local gauge symmetry (A.8) as
\begin{eqnarray} 
\psi^{\prime}_{n}(\vec{x},t)=e^{i\alpha_{n}(0)}
\psi_{n}(\vec{x},t)
\end{eqnarray}
{\em independently} of the value of $t$. 
This transformation is derived by using the exact 
representation (A.6) or (A.7), and it implies that 
$\psi_{n}(\vec{x},t)$ is a physical object since 
$\psi_{n}(\vec{x},t)$  stays in the same 
ray~\cite{dirac, streater} 
under an arbitrary hidden local gauge transformation.

\subsection { Adiabatic phase}
 We choose the specific basis set~\cite{berry}
\begin{eqnarray}
&&\hat{H}(\frac{\hbar}{i}\frac{\partial}{\partial\vec{x}},
\vec{x},X(t))v_{n}(\vec{x},t)
={\cal E}_{n}(X(t))v_{n}(\vec{x},t), 
\end{eqnarray}
and by assuming the dominance of diagonal elements in (A.7) in 
the adiabatic approximation, we have
\begin{eqnarray}
&&\psi_{n}(\vec{x},t)\\
&&\simeq v_{n}(\vec{x},t)
\exp\{-\frac{i}{\hbar}\int_{0}^{t}[{\cal E}_{n}(X(t))
-\langle n|i\hbar\frac{\partial}{\partial t}|n\rangle]dt\}.
\nonumber
\end{eqnarray}
which is confirmed to be invariant under the hidden local 
symmetry up to a constant phase. The product
\begin{eqnarray}
&&\psi_{n}(\vec{x},0)^{\dagger}\psi_{n}(\vec{x},T)
\nonumber\\
&&=v_{n}(0,\vec{x})^{\dagger}v_{n}(T,\vec{x})
\nonumber\\
&&\times\exp\{-\frac{i}{\hbar}\int_{0}^{T}[{\cal E}_{n}(X(t))
-\langle n|i\hbar\frac{\partial}{\partial t}|n\rangle]dt\}.
\end{eqnarray}
is thus manifestly invariant under the hidden local symmetry. 
By choosing the hidden local gauge such that 
$v_{n}(T,\vec{x})=v_{n}(0,\vec{x})$, the pre-factor 
$v_{n}(0,\vec{x})^{\dagger}v_{n}(T,\vec{x})$ becomes 
real and positive, and the phase factor in (A.12) defines a 
physical quantity uniquely. After this gauge fixing, the phase 
in (A.12) is still invariant under residual gauge transformations
 satisfying the periodic boundary condition 
$\alpha_{n}(0)=\alpha_{n}(T)$.   

\subsection{Non-adiabatic phase}

We start with the basic 
assumptions~\cite{aharonov,samuel}
\begin{eqnarray} 
&&i\hbar\partial_{t}\psi(t,\vec{x})=\hat{H}(t)\psi(t,\vec{x}),
\ \ \ 
\int d^{3}x \psi^{\dagger}(t,\vec{x})\psi(t,\vec{x})=1,
\nonumber\\
&&\psi(t,\vec{x})=e^{i\phi(t)}\tilde{\psi}(t,\vec{x}),\ \ \ \ 
\tilde{\psi}(T,\vec{x})=\tilde{\psi}(0,\vec{x}),\nonumber\\
&&\phi(T)=\phi, \ \ \ \ \phi(0)=0.
\end{eqnarray} 
We now choose the first element of the 
complete orthonormal set $\{v_{n}(t,\vec{x}) \}$ in (A.1) such 
that  
\begin{eqnarray}
v_{1}(t,\vec{x})=\tilde{\psi}(t,\vec{x}),
\end{eqnarray}
which is possible since $\{v_{m}(t,\vec{x}) \}$ is an 
arbitrary complete orthonormal set.
The  amplitude $\psi_{1}(t,\vec{x})$ in (A.7) satisfies the 
Schr\"{o}dinger equation and 
\begin{eqnarray}
\psi_{1}(0,\vec{x})=v_{1}(0,\vec{x})=\psi(0,\vec{x}).
\end{eqnarray}
We thus have~\cite{fujikawa3}
\begin{eqnarray}
\psi(t,\vec{x})&=&\psi_{1}(t,\vec{x})\nonumber\\
&=&v_{1}(t,\vec{x})\exp\{-\frac{i}{\hbar}
[\int_{0}^{t}dt\int d^{3}x v^{\dagger}_{1}(t,\vec{x})
\hat{H}v_{1}(t,\vec{x})\nonumber\\
&&\hspace{1.5 cm}-\int_{0}^{t}dt\int d^{3}x 
v^{\dagger}_{1}(t,\vec{x})i\hbar\partial_{t}v_{1}(t,\vec{x})]\}
\end{eqnarray}
where the last structure is fixed by noting 
$\psi(t,\vec{x})=v_{1}(t,\vec{x})e^{i\phi(t)}$ by 
{\em assumtion} (A.13), namely, by the assumption that only
the diagonal component survives for $\psi_{1}(t,\vec{x})$ in 
(A.7). 

 The amplitude $\psi(t,\vec{x})$ 
is invariant under the hidden local symmetry 
$v_{1}(t,\vec{x})\rightarrow e^{i\alpha_{1}(t)}v_{1}(t,\vec{x})$
up to a constant phase, $\psi(t,\vec{x})\rightarrow 
e^{i\alpha_{1}(0)}\psi(t,\vec{x})$. 
The quantity 
\begin{eqnarray}
&&\psi^{\dagger}(0,\vec{x})\psi(T,\vec{x})\nonumber\\
&&=v_{1}^{\dagger}(0,\vec{x})v_{1}(T,\vec{x})
\exp\{-\frac{i}{\hbar}
[\int_{0}^{T}dt\int d^{3}x v^{\dagger}_{1}(t,\vec{x})
\hat{H}v_{1}(t,\vec{x})\nonumber\\
&&\hspace{1.5 cm}-\int_{0}^{T}dt\int d^{3}x 
v^{\dagger}_{1}(t,\vec{x})i\hbar\partial_{t}v_{1}(t,\vec{x})]\}
\end{eqnarray}
is thus manifestly invariant under the hidden local symmetry
with a fixed Hamiltonian.  
If one chooses the gauge such that
$v_{1}(0,\vec{x})=v_{1}(T,\vec{x})$ as in our starting 
construction (A.1), the exponential factor in (A.17) extracts 
the entire phase from the gauge invariant quantity and , in 
particular, the non-adiabatic phase~\cite{aharonov} is given by
\begin{eqnarray}
\beta=\oint dt\int d^{3}x 
v^{\dagger}_{1}(t,\vec{x})i\hbar\partial_{t}v_{1}(t,\vec{x}).
\end{eqnarray}
It is clear that the geometric term in 
(A.17) is determined uniquely by the hidden local symmetry in 
(A.8) without referring to any explicit form of the Hamiltonian.
 The basis set $\{v_{n}(t,\vec{x})\}$ specify 
the coordinates in the functional space, and they do not satisfy
the Schr\"{o}dinger equation nor are the eigenvectors of 
$\hat{H}$ in general. 

\bibliography{apssamp}

\end{document}